\documentstyle[12pt]{article}
\font\titlefont=cmbx10 scaled \magstep3
\begin{document}

\begin{center} {\titlefont Preparing the State of a Black
Hole}\footnote{Dedicated to Mario Novello on the occasion of
his sixtieth birthday.}
\\
 
\vskip .3in Jacob D. BEKENSTEIN
\vskip .1in The Racah Institute of Physics, Hebrew University
of Jerusalem\\ Givat Ram, Jerusalem 91904, Israel\\
\end{center}

\vskip .2in

\begin{abstract}  Measurements of the mass or angular momentum
of a black hole are onerous, particularly if they have to be
frequently repeated, as when one is required to transform a
black hole to prescribed parameters. Irradiating a black hole
of the Kerr-Newman family with scalar or electromagnetic waves
provides a way to drive it to prescribed values of its mass,
charge and angular momentum {\it without\/} the need to
repeatedly  measure mass or angular momentum throughout the
process. I describe the mechanism, which is based on
Zel'dovich-Misner superradiance and its analog for charged
black holes.  It represents a possible step in the development
of preparation procedures for quantum black holes.
\end{abstract}

\baselineskip=14pt

\section{Introduction}  In general relativity a classical black
hole in equilibrium is singled out by just a few  parameters. 
The more mundane are mass $M$, electric charge $Q$ and angular
momentum vector ${\bf J}$ in which case we have a Kerr-Newman
black hole \cite{MTW}.  One could add parameters: magnetic
monopole, skyrmion number \cite{droz} and the like, but I shall
confine myself to the short list in order to make the story
clear.  Much of classical black hole physics consists of
considering processes and phenomena in connection with a
particular Kerr-Newman black hole or sequence of Kerr-Newman
black holes.  Little thought is given as to how to produce a
black hole with sharp values of $M$, $Q$ and
${\bf J}$.  This is no trivial task.  For example, total
gravitational collapse, accompanied as it probably always is by
matter ejection, radiation, etc., is apt to yield a black hole
whose parameters are not well known.  It would seem that to
prepare a black hole with sharp values of $M$, $Q$ and
${\bf J}$ one would have to follow the collapse by various
transformations of the hole while continuously keeping track of
the values of these parameters.  

Whereas measuring $Q$ of a black hole is relatively
straightforward,  measuring $M$ or ${\bf J}$ is no small matter
!  Mass is measurable only gravitationally, basically by keeping
track of a neutral test particle moving under the influence of
the hole.  ${\bf J}$ can also be deduced from the hole's
influence on a neutral test  particle.  However, the metric
component through which ${\bf J}$ primarily expresses itself,
the same that gives rise to the Lense-Thirring effect
\cite{MTW}, is relatively weak and rapidly decreasing with
distance from the hole (the effect is just becoming accessible
to modern space technology).  So the determination of ${\bf J}$
is even harder than that of $M$. Of course, ${\bf J}$ can also
be determined by first measuring the black hole's magnetic
dipole moment from its effects on a {\it charged\/} test
particle.  However, such a moment exists only for a charged
black hole, and to infer ${\bf J}$ from the measurement, both
$M$ and $Q$ (or at least their ratio
\cite{MTW}) must also be known.  Since a series of measurements
would seem to be be required to prepare a black hole state with
prescribed parameters, one would seem to be faced here with the
need for a sequence of difficult multiple measurements.

The problem of preparing a black hole state is compounded in
the quantum domain, Quantum Kerr black holes should have
available states with definite mass, charge and squared angular
momentum, as well as definite $z$ component of angular momentum
\cite{ann_arbor}; this is so because all these quantities are
represented in usual physics by commuting hermitian operators. 
In quantum theory of ordinary systems one measures an
observable $\hat O$, and if the measurement gives the
eigenvalue $o_i$ of $\hat O$, then one knows the system has
been prepared in the eigenstate of $\hat O$ whose eigenvalue is
$o_i$ (if $o_i$ is a degenerate eigenvalue we need more
observables to specify the prepared state).  How do we prepare
a quantum black hole state in this sense ?    Since, as
discussed, the  black hole observables are not so easily
measured even classically, we may enquire whether there is a
better way to prepare a quantum black hole state ? 

As a preparation for this challenge I discuss in this paper a
method for preparing a classical Kerr-Newman black hole in a
prescribed state, a method which requires only
one---initial---set of measurements, as opposed to a sequence
of measurements. In pedagogical spirit  I first set out the
method as applied to spherical black holes (Sec.~\ref{sec:RN}),
then describe the necessary modifications for treating rotating
black holes (Sec.~\ref{sec:K}), and only then join all parts
for the treatment of generic Kerr-Newman black holes
(Sec.~\ref{sec:KN}). Although my treatment is classical, I also
discuss (Sec.~\ref{sec:Hawking})  how to overcome difficulties
for the method arising from Hawking radiation.  

Unless otherwise stated, I work in units with $G=c=1$.

\section{Preparing a  Reissner-Nordstr\"om state}
\label{sec:RN}

I first explain the basic ideas in the simple context of
spherical black holes.   Consider a Reissner-Nordstr\"om black
hole (${\bf J}=0$) of mass $M$ and charge $Q$  with $|Q|\leq
M$.  The event horizon area of the hole is 
\begin{equation}  A={\cal A}(M,Q)\equiv 4\pi (M+\sqrt{M^2-Q^2}\
)^2
\label{A}
\end{equation}  Taking the total differential we have
\begin{eqnarray}
\Theta_{RN}\ d A&=&d M-\Phi d Q
\label{dA}
\\
\Theta_{\rm RN} &\equiv& {\scriptstyle 1\over \scriptstyle 2}
\sqrt{M^2-Q^2}\ A^{-1}
\label{ThetaRN}
\\
\Phi&\equiv& (4\pi/A)^{1/2}Q . 
\label{Phi}
\end{eqnarray}Here $\Phi$ is the electric potential of the
black hole ($-1<\Phi<+1$).  

Our tool for transforming the black hole will be a charged
Klein-Gordon field coupled to electromagnetism with charge
$\varepsilon$.   The question of the field's rest mass $m_s$
becomes relevant because, as will become clear, we would like
the energy
$\hbar\omega$ of the corresponding quanta---which must exceed
$m_s$---to be able to take up all values in the range
$[0,|\varepsilon\Phi|]$.  Although the known scalar fields of
these kind, e.g. the charged
$\pi$ field, are popularly construed as being quite massive, in
our units $m_\pi\sim 10^{-16}|\varepsilon_\pi|$. Thus I foresee
no problems in covering the desired range of
$\hbar\omega$ except for Reissner-Nordstr\"om black holes
extremely close to the Schwarzschild limit (say $|\Phi|\ll
10^{-13}$).  

We shall assume that surrounding the black hole at large
distance is a device---the {\it radiator\/} for short---capable
of irradiating the hole with coherent waves of this (bosonic)
scalar field.  Coherent waves are the same solutions of the
scalar field equation that we would use as modes in the quantum
treatment of the field.  In addition we shall need a {\it
monitor\/} that keeps track of the intensity of the incident
and scattered waves.  

Let us suppose a train of {\it spherical\/} wave modes of this
sort with (positive) frequency $\omega$ and $\varepsilon$ of
the {\it same sign\/} as $Q$ converge concentrically onto the
hole, and are partially absorbed and partially reflected off
it. Now if the hole absorbs a total of
${\cal N}$ quanta from the waves, the remainder of which is
scattered back, then  obviously the hole's parameters undergo
changes
$\delta M={\cal N}\hbar\omega$ and $\delta Q={\cal
N}\varepsilon$.  According to Eq.~(\ref{dA}), 
\begin{equation}
\delta A= \hbar {\cal N}\Theta_{\rm
RN}{}^{-1}(\omega-\varepsilon\Phi /\hbar).
\label{dA1}
\end{equation} Naively we would assume ${\cal N} > 0$.  Indeed,
if $\omega>\varepsilon\Phi /\hbar$ we then have from this
equation that $\delta A>0$ in harmony with Hawking's area
theorem \cite{hawking_area}. However, nobody can stop us from
taking $\omega<\varepsilon\Phi /\hbar$.  In fact, our tacit
assumption that $m_s\ll |\varepsilon|$ is tantamount to
allowing $\hbar\omega\ll |\varepsilon|$.  Since for the
Reissner-Nordstr\" om black hole $|\Phi|\leq 1$, we can in fact
arrange for
$\omega<\varepsilon\Phi /\hbar$, unless, of course, the hole is
very close to Schwarzschild.  Since we can make
${\cal N}$ large (by taking the incident wave strong) the
process is classical, yet Eq.~(\ref{dA1}) seems to suggest that
$\delta A<0$ in contradiction with Hawking's theorem.  One is
thus forced to accept that when $\omega<\varepsilon\Phi /\hbar$
then ${\cal N}<0$, i.e., the scattering {\it amplifies\/} the
wave \cite{bek73}.  This is the electromagnetic counterpart of
Zel'dovich-Misner superradiance from a Kerr black hole
\cite{zeldovich,misner}.  All this means that the reflection
coefficient ${\cal R}_{\rm RN}$ for the scalar waves off the
black hole in the vicinity of the {\it transition point\/}
$\omega=\varepsilon\Phi /\hbar$ may be expanded in a Taylor
series of the form
\cite{bekschif}
\begin{equation} {\cal R}_{\rm RN}=1-\alpha_{\rm RN}\cdot
(\omega-\varepsilon\Phi /\hbar)+O\left((\omega-\varepsilon\Phi
/\hbar)^3\right)
\label{aRN}
\end{equation} with some
$\alpha_{\rm RN}(\varepsilon,M,Q)>0$.     

If $\omega$ is tuned exactly to $\varepsilon\Phi /\hbar$, our
result tells us that the wave's intensity will not be affected
upon scattering---although the wave may thereby acquire a phase
shift.  Thus neither $M$ nor
$Q$ will change. This equilibrium between hole and
waves at the transition point $\omega=\varepsilon\Phi /\hbar$ is
a stable one.  For suppose the black hole parameters are
sightly shifted by an external agency from the transition point
so that $\omega$ slightly exceeds $\varepsilon\Phi/\hbar$. 
This is the normal regime and the black hole will absorb
radiation:  $\delta Q={\cal N}\varepsilon$ and $\delta M={\cal
N}\hbar\omega$ with ${\cal N}>0$.  Then from Eq.~(\ref{Phi}) we
find that this results in the change $\delta\Phi={\cal
N}\varepsilon(M+(M^2-Q^2)^{1/2})^{-1}$, so that
$\varepsilon\delta\Phi>0$.  Thus the black hole evolves back in
the direction of the transition point
$\omega=\varepsilon\Phi/\hbar$, and cannot stop until it
gets there.   Similarly, if the black hole is perturbed into the
superradiant regime (${\cal N}<0$),
$\varepsilon\delta\Phi<0$ and again the hole evolves back to the
transition point.  Thus the Reissner-Nordstr\" om transition
point is an attractor.

We can exploit this stability to prepare a Reissner-Nordstr\"
om hole with any prescribed parameters,  $\bar M$ and $\bar
Q$.  Our starting point can be any Reissner-Nordstr\" om black
hole whose horizon area $A$ lies below $\bar A\equiv {\cal
A}(\bar M,\bar Q)$.  For if  $A$ were bigger, we could never
reach the desired state classically---by Hawking's area
theorem---but would require for this the assistance of some
quantum effect, like Hawking radiation, which acts very slowly
for large black holes.  So how do we determine $A$ without
having to measure $M$ ?

This can be done by exposing the hole to a (very weak) test
charged scalar wave of the type mentioned earlier.  The idea is
to use this as a probe of the black hole state which has
negligible back effect.  By sweeping the $\omega$ of the wave
back and forth and having the monitor check the wave's
intensity gain upon scattering, we can zero in on the
transition point $\omega=\varepsilon\Phi/\hbar$ without
significantly affecting any of the hole's parameters.  Thus we
can determine $\Phi$ {\it directly\/}.  (In case the sign of
$\varepsilon$ was chosen infelicitously, zeroing-in is
impossible, and the opposite sign must be selected.)  Now
according to Eq.~(\ref{Phi}), if we can also find
$Q$, we know
$A$.  Measuring $Q$ involves no complications.  We may, for
example, measure the electric field at a couple of points far
away from the hole along the radial direction, and fit a
Coulomb law to the measurements  to isolate the value of the
initial charge,
$Q_0$.  This is much easier than measuring the gravitational
field because actual charged particles, which we would use as
probes in both types of measurements, have a large
charge-to-mass ratio in our units ($10^{18}$ for the
electron).  Thus we have a way to check if the initial horizon
area,  $A_0$, complies with the requirements from Hawking's
theorem.  And from Eq.~(\ref{A}), we can determine the initial
mass:
\begin{equation} M_0=(A_0/16\pi)^{1/2}(1+4\pi Q_0{}^2/ A_0)
\label{initmass}
\end{equation}

We next cause $A$ to increase at constant charge $Q_0$ until it
reaches ${\bar A}$.  This is easily done, for example, by
gradually dropping {\it neutral\/} matter with {\it zero\/}
angular momentum into the hole; according to Eq.~(\ref{dA}), at
constant charge we indeed need to increment the hole's mass in
order to increase its horizon area.  From Eq.~(\ref{A}) we know
that the final mass in this process must be (recall that
$Q=Q_0$ throughout the process)
\begin{equation} M_1=(\bar A/16\pi)^{1/2}(1+4\pi Q_0{}^2/ \bar
A).  
\end{equation} Comparison of this with Eq.~(\ref{initmass})
will tell us how much energy to add to the hole to reach $\bar
A$.

The process just described will in general change $\Phi$ (but
no its sign) and shift the transition point.  But we can again
lock the hole to the new transition point by repeating the
original $\omega$-sweeping procedure.  Once the hole is at the
transition point, we increase the intensity of the waves and
{\it slightly\/} shift the said $\omega$ in the direction of
$\varepsilon\bar\Phi/\hbar$, where $\bar\Phi$ is computed from
$\bar Q$ and $\bar M$ by means of Eq.~(\ref{Phi}).  By the
stability condition we know that the black hole will follow
suit and change its parameters so that its instantaneous
$\Phi$ evolves in the direction of the desired $\bar\Phi$. 
Thus by slowly shifting $\omega$ in the same sense, we drag the
actual $\Phi$ in the desired direction.  (If the initial $Q$ is
opposite in sign to $\bar Q$, we shall have to switch the sign
of $\varepsilon$ as $\Phi$ passes through zero.) The question
is, are we changing $A$ by this procedure ?

A look at Eq.~(\ref{dA1}) informs us that the change in $A$ in
the process amounts to  $\delta A=\Theta_{\rm
RN}{}^{-1}\cdot(1-\varepsilon\Phi/\hbar\omega)\cdot\delta M$. 
Thus if the sweep of $\omega$ is carried out sufficiently slowly
so that the hole stays very close to the transition point (
$|1-\varepsilon\Phi/\hbar\omega|\ll 1$), the concomitant change
in $A$ will be entirely negligible on the scale of the changes
of $M$ and $Q$.  This is just another example of an adiabatic
black hole process which leaves the horizon area unchanged
\cite{bek98,mayo98,duez}.  When $\omega$ reaches the value
$\varepsilon\bar\Phi/\hbar$, we have managed to bring the hole
to the potential $\bar\Phi$ and area $\bar A$.  Because
Eqs.~(\ref{A}) and (\ref{Phi}) can be solved uniquely for $M$
and $Q$, we have obtained the black hole with the prescribed
mass $\bar M$ and charge $\bar Q$.
   
\section{Preparing a  Kerr state}
\label{sec:K}

For a Kerr black hole ($Q=0$) with mass $M$ and angular momentum
$J$ about a specified axis ($J \leq M^2$),  the analogs of
Eqs.~(\ref{A})-(\ref{Phi}) are
\begin{eqnarray} A={\cal A}(M,J)&=&4\pi [r_+{}^2+(J/M)^2]
\label{AK}
\\
\Theta_{\rm K}\ d A &=& d M - \Omega\, d J
\label{dAK}
\\
\Theta_{\rm K} &\equiv& {\scriptstyle 1\over \scriptstyle 2}
\sqrt{M^2-(J/M)^2}\ A^{-1}
\label{ThetaK}
\\
\Omega &\equiv& 4\pi (J/MA)
\label{Omega}
\end{eqnarray}    Here  $r_+\equiv M+\sqrt{M^2-J^2/M^2}$ and
$\Omega$ is the hole's rotational angular frequency
($-(2M)^{-1}  < \Omega< +(2M)^{-1} $).  Henceforth I will
identify the hole's rotation axis with the $z$-axis.   More
appropriate here than the charged scalar field of
Sec.~\ref{sec:RN} is the electromagnetic---hence also
bosonic---field.  It is more suitable because its masslessness
guarantees the existence of photons with energies $\hbar\omega$
as low as desired.  In the sequel we shall have need of quanta
frequencies which fall below $|\Omega|$ times a small integer,
and for macroscopic black holes this means a very low
$\omega$.   

Now focus on an electromagnetic mode of frequency $\omega$ with
the asymptotic angular form of a vector spherical harmonic 
${\cal Y}_{j \mu}(\theta,\phi)$ (with the azimuthal angle
$\phi$ and the integer $\mu$---the $z$ component of angular
momentum---both referred to the $z$ axis).  Suppose that when
the radiator irradiates the black hole with a train of such
wavemodes, the hole absorbs
${\cal N}$ quanta.  Then $\delta M={\cal N}\hbar\omega$ while
$\delta J= {\cal N}\hbar \mu$, and it can be seen from
Eq.~(\ref{dAK}) that
\begin{equation}
\delta A= \hbar {\cal N}\Theta_{\rm K}{}^{-1}(\omega-\mu
\Omega).
\label{dA2}
\end{equation} In the regime $\omega>\mu\Omega$ the assumption
that ${\cal N}>0$ implies that the horizon area  increases, in
harmony with Hawking's theorem.  However, for
$\omega<\mu\Omega$ there will be a conflict with the theorem
unless we assume that ${\cal N}<0$, i.e., that the black hole
reinforces the wave.  This is the original Zel'dovich-Misner
superradiance.  In analogy with Eq.~(\ref{aRN}) we may infer
that the reflection coefficient in the vicinity of the
transition point $\omega=\mu\Omega$ looks like
\begin{equation} {\cal R}_{\rm K}=1-\alpha_{\rm K}\cdot
(\omega-\mu\Omega)+O\left((\omega-\mu\Omega)^3\right)
\label{aK}
\end{equation}  with $\alpha_{\rm K}(\mu,M,J)>0$.          

Again, if we tune $\omega$ to $\mu\Omega$, the wave is neither
amplified nor depleted: equilibrium of the hole with the waves
ensues at the transition point.  This, again, is a stable
equilibrium.  For if we push the hole slightly off the
transition point so that $\omega$ exceeds
$\mu\Omega$, the hole will absorb: $\delta M={\cal N}\hbar
\omega$ and $\delta J= {\cal N}\hbar
\mu$ with ${\cal N}>0$.  From  Eq.~(\ref{Omega}) we find that
the consequent  $\mu\delta\Omega$ is a sum of two
terms, both positive by virtue of the constraint $J\leq M^2$. 
The absorption thus  acts to bring the hole back to the
transition point  $\omega=\mu\Omega$.  Similarly, if we perturb
the hole into the superradiant regime, the corresponding 
 $\mu\delta\Omega$ will be {\it negative\/}, which again
tends to bring the hole back to the transition point. 
Therefore, the Kerr transition point is also an attractor.  
  
Of course, we have only verified stability at the transition
point against perturbations which preserve the $z$ axis.  What
if we tilt this axis slightly from the axis set by the radiator
(with respect to which the mode
${\cal Y}_{j \mu}(\theta,\phi)$ is determined) ?  Such a slight
tilt would mean that the waves falling on the hole would be
mostly in the primary mode ${\cal Y}_{j \mu}(\theta,\phi)$ but
with a slight admixture of various other values of
$\mu$. Some of these secondary modes might actually
 superradiate while the primary mode acts normally, or
viceversa.  There is no question, however, of this new feature
overturning our mentioned conclusion.  The very  weakness of
the secondary modes would cause their effects on the hole to be
swamped by that of the primary; the convergence
$\mu\Omega\rightarrow \omega$ would still take place.  And
because the hole is gaining or loosing angular momentum about
the direction imposed by the radiator, it is intuitively clear
that the black hole's rotation axis $z$ will gradually line up
with the primary mode's axis.  The Kerr black hole at the
transition point is stable overall.   

So suppose we are given a Schwarzschild or a Kerr black hole and
required to prepare it in a Kerr state with prescribed
parameters $\bar M$ and $\bar J$ about a prescribed axis.  As
in the previous example, this can be done classically only if
the initial horizon's area is below $\bar A={\cal A}(\bar M,\bar
J)$.  I have not hit upon an appropriate analog of the method
of Sec.~\ref{sec:RN} to determine the area in the pure Kerr (or
pure Schwarzschild) case.  However, it stands to reason that
the initial  $A$ can be determined from the scattering
crossection of the hole to very weak incident {\it plane\/}
waves.  I will not go into the details of such a method, but
just assume they can be worked out. By using this procedure
once only we can determine the initial horizon area, $A_0$, of
the black hole.  If the area condition is satisfied, we proceed
as follows.

First we cause the radiator to irradiate the hole with a (very
weak) test wave which is asymptotically of ${\cal Y}_{j
\mu}(\theta,\phi)$ form (as defined by the arbitrarily selected
radiator axis), and let its frequency sweep across the spectrum
while the said axis is allowed to range over the sphere.  All
this while the monitor checks the intensity of the scattered
wave.  We can thus home in on the transition point
$\omega=\mu\Omega$ and on the axis of the hole---the $z$ axis. 
This is the combination for which the amplification factor
$a_{\rm K}$ for the wavemode in question is unity.  (If the
sign of $\mu$ has been chosen inappropriately, homing-in is
impossible, and the opposite sign must be selected.)  The
process just described  does not increase
$A$ significantly due to the test wave's weakness.  

Thus we have determined the initial horizon area, $A_0$, axis
direction and the angular frequency,
$\Omega_0$, of the hole.  And from Eqs.~(\ref{Omega})
and~(\ref{AK}) we can determine the initial mass
\begin{equation}
M_0=(A_0/16\pi)^{1/2}(1-A_0\Omega_0^2/4\pi)^{-1/2}
\end{equation} from which follows the initial angular momentum
$J_0=(M_0A_0\Omega_0/4\pi)$. Next we cause $A$ to grow to $\bar
A$ by gradually adding  {\it neutral\/} matter or waves with
{\it zero\/} angular momentum to the hole; according to
Eq.~(\ref{dAK}), to raise
$A$ under this condition we indeed need to make $M$ larger.  
And from Eq.~(\ref{AK}) we see that the final mass of the
process (recall that $J$ is fixed at $J_0$) is given by the
Christodoulou-style formula \cite{Christodoulou}
\begin{equation} M_1=(\bar A/16\pi)^{1/2}(1+64\pi^2
J_0{}^2/\bar A^2)^{1/2}
\end{equation} The difference $M_1-M_0$ is the energy that must
be added to raise the horizon area to $\bar A$.

 Suppose now that the determined $z$ axis (which was unaffected
by the step just carried out) is collinear with the prescribed
rotation axis.  Then as in the Reissner-Nordstr\" om case, we
may now ``drag'' the  parameter $\Omega$ from $\Omega_0$ to the
$\bar\Omega$ corresponding to $\bar M$ and
$\bar{\bf J}$ by the simple expedient of consistently shifting
$\omega$ slightly away from the instantaneous transition point
in the direction of $\mu\bar\Omega$.  (Of course, in case
$\Omega_0$ and
$\bar \Omega$ are opposite in sign, it will be necessary to
switch the sign of $\mu$ as $\Omega$ passes through zero.) If,
on the contrary, the determined $z$ axis proves not collinear
with the prescribed rotation axis, we first drag
$\Omega$ to zero converting the hole into a Schwarzschild one. 
Whether the original black hole was Schwarzschild or Kerr, we
may now, by irradiating the hole with a wave of angular form
${\cal Y}_{j
\mu}(\theta,\phi)$ with
$\mu$ defined with respect to the prescribed axis, drag
$\Omega$ from zero to $\bar\Omega$.  Of course the final black
hole's rotation will be about the prescribed axis. According to
Eq.~(\ref{dA2}), the increment in area
$A$ during the two processes in this paragraph is
$\Theta_{\rm K}{}^{-1}\cdot (1-\mu\Omega/\omega)\cdot\delta
M$.  Thus by changing $\omega$ sufficiently slowly so that
$\mu\Omega$ keeps pace with it, we can accomplish the
transformation with as small a change of $A$ as desired.  Since
$A$ had already reached the desired value
$\bar A$, and since Eqs.~(\ref{AK}) and (\ref{Omega}) can be
solved uniquely for $M$ and $J$ in terms of $A$ and $\Omega$,
we have reached the prescribed values $\bar M$ and
$\bar J$.  

\section{Preparing a generic Kerr-Newman state}
\label{sec:KN}

The generic stationary---Kerr-Newman---black hole has
parameters $M$, $Q$ and $J$ subject to the condition $M\geq
(Q^2/2+\sqrt{Q^4/4+J^2}\ )^{1/2}$.  The analogs of
Eqs.~(\ref{Phi}), (\ref{AK}) and (\ref{Omega})  are here
\cite{MTW}
\begin{eqnarray} A={\cal A}(M,J,Q)&=&4\pi [r_+{}^2+(J/M)^2]
\label{AKN}
\\
\Theta_{\rm KN}\ d A &=& d M - \Omega\, d J-\Phi d Q
\label{dAKN}
\\
\Theta_{\rm KN} &\equiv& {\scriptstyle 1\over \scriptstyle 2}
\sqrt{M^2-Q^2-(J/M)^2}\ A^{-1}
\label{ThetaKN}
\\
\Omega &\equiv& 4\pi (J/MA)
\label{Omega3}
\\
\Phi&\equiv& 4\pi Qr_+/A  
\label{Phi3}
\end{eqnarray} with $r_+\equiv
M+\sqrt{M^2-Q^2-(J/M)^2}$.        

Let us unify the discussions of Secs.~\ref{sec:RN} and
\ref{sec:K} by considering {\it charged\/} wavemodes with
definite angular momentum.  These could refer to either a
scalar or a vector field since both are bosonic. We suppose that
at the outset the axis with respect to which $\mu$ is defined
coincides with the hole's rotation axis---the $z$-axis. The
analysis is then more complicated than before, but it should
come as no surprise, in view of the form of Eq.~(\ref{dAKN}),
that the transition point between normal and superradiant
regimes resides at
$\omega=\varepsilon\Phi/\hbar+\mu\Omega$. In analogy with
Eq.~(\ref{aRN}) and (\ref{aK}) we find the  reflection
coefficient in the vicinity of the transition point to be
\begin{equation} {\cal R}_{\rm KN}=1-\alpha_{\rm KN}\cdot
(\omega-\varepsilon\Phi/\hbar-\mu\Omega)+O\left((\omega-
\varepsilon\Phi/\hbar-\mu\Omega)^3\right)
\label{aKN}
\end{equation}   with some $\alpha_{\rm KN}(\varepsilon,\mu,
M,Q,J)>0$.  
 
Suppose we perturb the black hole so that it shifts into the
superradiant regime,
$\omega<\varepsilon\Phi/\hbar+\mu\Omega$.  This, of course,
leads to reinforcement of the wave at expense of the hole:
$\delta M={\cal N}\hbar\omega$, $\delta J={\cal N}\hbar\mu$ and
$\delta Q={\cal N}\varepsilon$ with
${\cal N}<0$ this time.  It takes a laborious calculation (I
used {\it Mathematica\/} to do the algebra) to show that, {\it
whatever\/} the ratio
$\varepsilon/\mu$, these increments give
$\varepsilon\delta\Phi/\hbar+\mu\delta\Omega<0$.  Thus the black
hole changes in such a sense as to approach the
transition point $\omega=\mu\Omega + \Phi\varepsilon/\hbar$. 
Likewise, if we push the hole into the normal regime, it will
absorb, $\varepsilon\delta\Phi/\hbar+\mu\delta\Omega>0$, and as
a consequence it will again be driven towards the transition
point.  The Kerr-Newman transition point is thus also an
attractor.  Just as in Sec.~\ref{sec:K} we can argue that this
property is  maintained in the face of perturbations of the
hole's rotation axis.

How to prepare a state with prescribed parameters $\bar M$,
$\bar Q$ and $\bar J$ starting from a given Kerr-Newman hole ? 
As previously, we first have to check that $A\leq \bar A\equiv
{\cal A}(\bar M,\bar Q,\bar J)$.  Following the example of
Sec.~\ref{sec:RN}, we cause the radiator to irradiate the black
hole with {\it two separate\/} wavemode trains, one of
asymptotically spherical ($\mu=0$) scalar modes with charge
$\varepsilon$, and the second of electromagnetic modes
($\varepsilon=0$)  all with the same $j$ and azimuthal index
$\mu\neq 0$.  The radiator is supposed to sweep the frequency
of the first mode,  through a broad range and, with help of the
monitor, tune it to the value
$\varepsilon\Phi/\hbar$ (as before, this requires the
appropriate choice for the sign of $\varepsilon$).  Likewise,
the radiator is charged with aligning its reference axis with
the hole's and tuning the frequency
$\omega_2$, of the second mode to
$\mu\Omega$ (again after a felicitous choice of sign of $\mu$). 
Thus are the initial values
$\Phi_0$ and
$\Omega_0$ determined.  We also determine $Q_0$ this once only
from its Coulomb field as in Sec.~\ref{sec:RN}.  Now by
substituting $r_+$ from Eq.~(\ref{Phi3}) and $J/M$ from
Eq.~(\ref{Omega3}) in the expression for ${\cal A}(M,Q,J)$,
Eq.~(\ref{AKN}), we have
\begin{equation} A={4\pi\over (\Phi/Q)^2+\Omega^2}.
\label{help}
\end{equation} Thus we can determine the initial area, $A_0$,
and obtain
$Q_0$ and $(J/M)_0$ into the bargain.  

If $A_0<{\bar A}\equiv {\cal A}(\bar M,\bar Q,\bar J)$, we
proceed to determine the initial mass, $M_0$, by eliminating
$r_+$ between its definition and Eq.~(\ref{Phi3}), replacing
$J_0/M_0$ from Eqs.~(\ref{Omega3}), and simplifying with help
of Eq.~(\ref{help}): 
\begin{equation} M_0={\scriptstyle 1\over
\scriptstyle 2}(Q_0/\Phi_0)+2\pi (Q_0{}^3/\Phi_0 A_0)
\end{equation} We follow this by gradual addition of  {\it
neutral\/} matter or waves with {\it zero\/} angular momentum
to the hole in order to raise its area to $\bar A$.  According
to Eq.~(\ref{dAKN}) this indeed requires an increase in $M$. 
Solving Eq.~(\ref{AKN}) for $M$ we get for the mass at the end
of the process (recall that $J_0$ and $Q_0$ are unchanged) the
Christodoulou-Ruffini style formula \cite{CR}  
\begin{equation} M_1 = \left[{\bar A\over 16\pi}\left(1+{4\pi
Q_0{}^2\over
\bar A}\right)^2 + {4\pi J_0{}^2\over \bar A}\right]^{1/2}
\end{equation} The quantity $M_1-M_0$ is the energy we are
required to add to the hole. 

Of course the envisaged process has caused $\Phi$ and $\Omega$
to drift away from the transition points.  We, therefore, again
sweep the frequencies of the two wavemodes to recover the two
corresponding transition points.  We then shift the two mode
frequencies away from the two transition points in the
directions of $\varepsilon\bar\Phi/\hbar$ and
$\mu\bar\Omega$, respectively. This has the effect of dragging
the black hole's rotational angular frequency and electric
potential to the prescribed values $\bar\Omega$ and
$\bar\Phi$.  (As before, if initial and prescribed values
differ in sign, we have to switch the sign of $\varepsilon$ or
$\mu$ when $\Phi$ or $\Omega$, respectively, pass through
zero.)    Again, slowness of the dragging guarantees that the
horizon's area does not increase significantly on the scale of
the overall changes in the black hole's parameters. At the end
of the envisaged process the desired $\bar M$, $\bar Q$ and
$\bar J$ will have been reached because, as easily checked from
the expressions (\ref{AKN}), (\ref{Omega3}) and (\ref{Phi3}),
particular values of $\bar A,
\bar\Omega$ and
$\bar\Phi$ correspond to a unique set of values for
$\bar M$, $\bar Q$ and $\bar J$.     

\section{Beating the Hawking radiance}
\label{sec:Hawking}

All the above discussions ignored Hawking radiance which poses
some problems for our method.  For instance, because it emerges
in a range of modes, it tends to cause the hole parameters to
drift away from any prescribed values, e.g. away from the
transition point if set there initially.  This destabilization
be counteracted if the radiation incident from the radiator is
sufficiently strong.   We now determine how strong it must be. 

It is well known that Hawking radiation from a Kerr-Newman
black hole is thermal---apart from a distortion in the mean
occupation numbers of the various modes.  Thus we can estimate
the rate, as measured with respect to global time, at which the
hole loses mass to the radiation by using the Stefan-Boltzmannn
law appropriate to Hawking's temperature $T_{\rm
H}=4\hbar\Theta_{\rm KN}$ and radiating area $A$:
\begin{equation} |\dot M|_{\rm H}\approx {\pi^2 A T_{\rm
H}{}^4\over 60
\hbar^3}={4\pi^2\hbar[M^2-Q^2-(J/M)^2]^2\over 15A^3 }\leq
{\hbar\over 240\pi M^2}
\label{dotMH} 
\end{equation}    This must be compared with $\hbar\omega$
times the rate at which quanta are absorbed from the radiator
wavemodes (if
$\omega- \varepsilon\Phi-\mu\Omega>0$) or added to them (if
$\omega- \varepsilon\Phi-\mu\Omega<0$), namely

\begin{equation} |\dot M|_{\rm rad}=|1-{\cal R_{\rm KN}}|\cdot
N\hbar\omega\cdot(\Delta\omega/2\pi)
\label{dotMrad}
\end{equation} in both cases.  Here $N$ is the occupation
number of the incoming modes while the factor
$(\Delta\omega/2\pi)$ is the usual rate of flow of modes in one
dimension---since we keep
$\mu$ and $j$ fixed---assuming that those modes span a
bandwidth $\Delta\omega$. Since the radiator operates near the
transition point we can use Eq.~(\ref{aKN}) to approximate
${\cal R}_{\rm RN}$.   Now $\alpha_{\rm KN}$ has dimensions of
length and it measures absorption by the hole, so it is
intuitively clear that it must be of order $\surd A$; for
orientation I take $\alpha_{\rm KN}=\surd A\geq 2\surd\pi M$.  
Thus  provided we take \begin{equation} N\gg
{\omega/\Delta\omega\over 240\surd\pi\
|1-\varepsilon\Phi/(\hbar\omega)-\mu\Omega/\omega|
\ (M\omega)^3},
\label{condition}
\end{equation} the rate of change of $M$ attributable to the
radiator will strongly dominate $|\dot M|_{\rm H}$.  The hole
will then be stabilized at the transition point  against
Hawking's radiance since the last causes relatively slow
changes in $M$, as well as in $Q$ and $J$.  

As an illustration let us consider a plain Kerr black hole
locked into its transition point by a train of electromagnetic 
wavemodes with definite $\mu\geq 1$.  If the quality parameter
$|1-\mu\Omega/\omega|$, which measures the hole's departure
from the transition point, is $10^{-4}$, the r.h.s. of
condition (\ref{condition}) is actually smaller than
$24 (M\mu\Omega)^{-3}(\omega/\Delta\, \omega)$.  Since the
quanta are bosons, the condition is thus easy to satisfy if we
do not insist on almost monochromatic incoming radiation
($\Delta\omega/\omega$ not very small compared to unity), and
if the black hole is not almost a Schwarzschild one ($M\Omega$
not very small compared to unity).  Stabilizing a generic Kerr
hole at its transition point against Hawking radiation drift is
thus relatively easy. 

The Unruh radiation  \cite{unruh}, which emerges spontaneously
in a range of superradiant modes, is quite similar to Hawking's
in this same respect.  Thus its destabilizing effects can be
suppressed as well. Neither radiation need cause problems
during the stage in which the frequency $\omega$ is swept to
locate the black hole's transition point because the sweep can
be made quickly so that little spontaneous radiation is emitted
in the interim.

However, as mentioned, the stage in which $\Omega$ or $\Phi$
are ``dragged'' has to be protracted, for otherwise the process
would fail to be adiabatic and the horizon area $A$ would
increase substantially.  We do not want this to happen because
the first step of our process has already set $A$ to its final
value.  But---so it would seem---if the dragging is performed
too slowly,  a lot of Hawking radiation will get emitted in the
interim with consequent {\it quantum\/} decrease of $A$.  This
last eventuality would complicate the procedure we have set
forth for preparing the black hole state. 

Fortunately the same condition (\ref{condition}), which
guarantees that Hawking's radiance does not destabilize the
hole from its transition point, insures that the radiance
generates negligible changes of the horizon area in the course
of dragging.  Let us calculate the change in the logarithm of
$A$ which comes from Hawking's radiance (subscript ``H'') in
terms of mass changes attributable to the radiator (subscript
``rad''):
\begin{equation} (\Delta  \ln A)_{\rm H} = \int{\dot A_{\rm
H}\over A}\, dt=
\int{\dot A_{\rm H}\over A
\dot M_{\rm rad}}\, dM_{\rm rad}
\label{deltalog}
\end{equation}  Now according to Eq.~(\ref{dAKN}), the fact that
Hawking's radiance carries away angular momentum (charge) of the
same sign as the hole's angular momentum (charge) implies 
(remember that $\dot M_{\rm H}$ and $\dot A_{\rm H}$ are both
negative)  that $|\dot A|_{\rm H}<\Theta_{\rm KN}{}^{-1}|\dot
M|_{\rm H}$.  Thus it follows from Eq.~(\ref{deltalog}) that 
\begin{equation} |\Delta \ln A|_{\rm H} < \int{|\dot M|_{\rm
H}\over \Theta_{\rm KN}\, A |\dot M_{\rm rad}|}\, |dM_{\rm rad}|
\end{equation} where we have exploited the fact that $\dot
M_{\rm rad}$ and
$dM_{\rm rad}$ are of like sign. We now substitute here from
Eq.~(\ref{dotMH}) the second form for
$|\dot M|_{\rm H}$ as well as $|\dot M|_{\rm rad}$ from
(\ref{dotMrad}) with the previously discussed value 
$\alpha_{\rm KN}\geq 2\surd\pi M$.  Now because by condition
(\ref{condition}) most of $M's$ change is attributable to the
radiator, $dM_{\rm rad}\approx dM$ and $\dot M_{\rm rad}\approx
\dot M$.  In view of Eqs.~(\ref{AKN}) and (\ref{ThetaKN}) we now
have
\begin{equation} |\Delta \ln A|_{\rm H} <\int
{\omega/\Delta\omega\over 120\surd\pi\
|1-\varepsilon\Phi/(\hbar\omega)-\mu\Omega/\omega|\ N
(M\omega)^3}\ |d\ln M| 
\end{equation} It then follows from condition (\ref{condition})
that  $|\Delta
\ln A|_{\rm H}\ll |\Delta \ln M|$, i.e., the fractional change
in $A$ which is attributable to the Hawking radiance is
negligible compared to the overall fractional change in $M$ (or
in the other hole parameters for that matter) during
``dragging''. As already mentioned a the end of
Secs.~\ref{sec:RN} and
\ref{sec:K}, the change in $A$ directly  attributable to the
radiator is also entirely negligible on this scale.  Thus one
can indeed carry out slow dragging of $\Omega$ and $\Phi$
without having the horizon area increase significantly.

\medskip {\bf Acknowledgments\hskip 6pt} This research is
supported by grant No. 129/00-1 of the Israel Science
Foundation.


\begin{thebibliography}{99}

\bibitem{MTW} Misner  C. W.,  Thorne K. S. and  Wheeler J. A.,
1973,  {\it Gravitation\/}. Freeman, San Francisco

\bibitem{droz}  Droz S.,  Heusler M. and  Straumann N. (1991),
Phys. Lett. B  {\bf 268}, 371

\bibitem{ann_arbor}Bekenstein, J.D. (2002), in  Duff, M. and
Liu,  J. T. (eds.), {\it 2001: A Spacetime Odyssey\/}. World
Scientific Publishing, Singapore. hep-th/0107045.

\bibitem{hawking_area}Hawking, S.W. (1971),  Phys. Rev.
Letters, {\bf 26}, 1344.

\bibitem{bek73}Bekenstein, J.D. (1973)  Phys. Rev. D, {\bf 7},
949.

\bibitem{zeldovich}Zel'dovich,  Ya. B. (1971), Zh. Eksp. Teor.
Fiz. Pis'ma {\bf 14}, 270; [JETP Letters (1971), {\bf 14}, 180
].

\bibitem{misner}Misner, C. W.  (1972), Phys. Rev. Letters {\bf
28}, 994.

\bibitem{bekschif}Bekenstein, J. D.  and Schiffer, M. (1998),
Physical Review D{\bf  58},  64014.

\bibitem{bek98}Bekenstein, J.D. (1998) in  Iyer, B.R. and 
Bhawal, B. (eds.),  {\it Black Holes, Gravitational  Radiation
and the Universe\/}. Kluwer, Dordrecht.

\bibitem{mayo98}Mayo, A.E. (1998),  Phys. Rev. D {\bf 58},
104007.

\bibitem{duez}Duez,  M.W. et. al (1999),  Phys. Rev. D, {\bf
60}, 104024.

\bibitem{Christodoulou} Christodoulou D. 1970, Phys. Rev.
Lett.  {\bf 25}, 1596.

\bibitem{CR} Christodoulou D. and  Ruffini R. (1971),  Phys.
Rev. D {\bf 4}, 3552.

\bibitem{hawking2}Hawking, S.  W. (1975), Commun. Math. Phys.  
{\bf 43}, 199.

\bibitem{unruh}Unruh,  W. (1974), Phys. Rev. D {\bf 10}, 3194.

\end{thebibliography}
\end{document}